\documentclass[reprint,showpacs,showkeys,amsmath,amssymb,aps]{revtex4-1}

\usepackage{graphicx}% Include figure files
\usepackage{dcolumn}% Align table columns on decimal point
\usepackage{bm}% bold math
\usepackage{hyperref}% add hypertext capabilities

\begin{document}

\title{\v Cerenkov radio pulses from electromagnetic showers in the time-domain}

\author{Jaime Alvarez-Mu\~niz}
\affiliation{%
Depto. de F\'\i sica de Part\'\i culas
\& Instituto Galego de F\'\i sica de Altas Enerx\'\i as,
Universidade de Santiago de Compostela, 15782 Santiago
de Compostela, Spain
}%

\author{Andr\'es Romero-Wolf}
\affiliation{%
Dept. of Physics and Astronomy, 
University of Hawaii at Manoa, Honolulu, HI 96822, USA
}%

\author{Enrique Zas}
\affiliation{%
Depto. de F\'\i sica de Part\'\i culas
\& Instituto Galego de F\'\i sica de Altas Enerx\'\i as,
Universidad de Santiago de Compostela, 15782 Santiago
de Compostela, Spain
}%

\begin{abstract}
The electric field of the \v Cerenkov radio pulse produced by a single charged particle track
in a dielectric medium is derived from first principles. An algorithm 
is developed to obtain the pulse in the time domain for numerical calculations.
The algorithm is implemented in a Monte Carlo simulation of electromagnetic showers
in dense media (specifically designed for coherent radio emission applications) as might be induced 
by interactions of ultra-high energy neutrinos. The coherent \v Cerenkov radio emission produced
by such showers is obtained simultaneously both in the time and frequency domains.
A consistency check performed by Fourier-transforming the pulse
in time and comparing it to the frequency spectrum obtained directly in the simulations
yields, as expected, fully consistent results. 
The reversal of the time structure inside the \v Cerenkov cone and the signs of the
corresponding pulses are addressed in detail.
The results, besides testing algorithms used for reference calculations in the frequency domain,
shed new light into the properties of the radio pulse in the time domain. 
The shape of the pulse in the time domain is directly related to the depth development of the excess charge in the
shower and its width to the observation angle with respect to the \v Cerenkov direction. 
This information can be of great practical importance for interpreting actual data. 
\end{abstract}

%95.85.Bh Radio, microwave ( >1 mm)
%95.85.Ry Neutrino, muon, pion, and other elementary particles; cosmic rays
%29.40.-n Radiation detectors


\maketitle

%%%%%%%%%%%%%%%%%%%%%%%%%%%%%%%%%%%%%%%%%%
\section{Introduction}

It was nearly 50 years ago that Askaryan proposed to detect high energy 
particles through the coherent pulse they emit as they interact in a dense 
medium~\cite{Askaryan62}. As secondary electrons, positrons and gamma rays are 
produced they develop electromagnetic showers in the medium which
acquire an excess negative charge, which Askaryan estimated to be of order
$10\%$ of the total number of electrons and positrons.
This is so in spite of the interactions being 
completely charge symmetric, because matter in the medium only contains 
electrons. M\o ller, Bhabha and Compton 
scattering of matter electrons, accelerate them into the shower while 
electron-positron annihilation and Bhabha scattering decelerate the shower positrons 
thus also contributing to the excess charge, a mechanism referred to as the Askaryan effect. 
A more accurate calculation of the Askaryan effect
indicated that the excess charge is actually $\sim 25\%$ of the
total number of electrons and positrons~\cite{ZHS92}. 
Such an excess charge develops a coherent electromagnetic 
pulse as it travels through a non absorptive dielectric medium. The coherent
part of the pulse is mainly due to the wavelength components which are large
compared to the shower width. 
The energy radiated in the coherent pulse 
scales with the square of the excess charge and hence with the square of the 
shower energy. 
Such scaling naturally makes the detection of coherent radio pulses an 
attractive and promising technique for the detection of ultra high energy 
particles, such as cosmic rays. 

Radio detection of air showers was extensively studied in the 60's and
70's~\cite{allan71}. % and also in recent years \cite{ARENA08}.   
The drive to detect high energy neutrinos in the late 80's turned back the 
attention onto radio pulses produced by them in dense media such as natural
ice~\cite{frichter96} or the regolith beneath the Moon's 
surface~\cite{zheleznykh}. 
The first full simulations of the Askaryan effect and the coherent pulses 
created in dense media were obtained in the early 90's~\cite{ZHS92,ZHS91}, which
allowed more quantitative calculations and experimental programs were soon 
after started to search for neutrinos with arrays of antennas at 
Antarctica~\cite{RICE98} or with radio telescopes from Earth~\cite{Parkes96}. 
The Askaryan effect was measured for the first time 
firing photon bunches into sand at SLAC in 2000~\cite{Saltzberg_SLAC_sand} - and later
in other dielectric media including ice 
\cite{Gorham_SLAC_salt,Miocinovic_SLAC_sand,Gorham_SLAC_ice} - and  
since then the field has received an enormous boost, strengthening 
previous initiatives using antennas buried in ice ~\cite{RICE03,RICElimits} and 
radiotelescopes~\cite{Parkes07}, and developing new ones such as a balloon 
flown antenna array~\cite{ANITAlite,ANITA_2009_limits,ANITAlong}, new radiotelescope 
searches~\cite{GLUElimits,Kalyazin,NuMoon,LUNASKA,RESUN} and new radio
measurements of air showers \cite{ARENA08}. 

The first calculation of the radio emission
from electromagnetic showers used a specifically
designed Monte Carlo simulation code - the ZHS code - to calculate coherent radio pulses in 
ice~\cite{ZHS91,ZHS92}. The code has been extended to include the LPM
effect~\cite{alz97}, to calculate in an approximate manner hadronic showers~\cite{alz98}
and neutrino-induced showers \cite{alvz99}, 
to treat other dielectric media~\cite{alvz06}, and to perform an optimal statistical thinning
that allows the simulation of pulses from ultrahigh energy 
showers~\cite{aljpz09}, and remains as a reference in the field. 
This code was designed to calculate the 
Fourier components of the electric field in the frequency domain. Alternative
simulations using other codes such as GEANT3 \cite{almvz03,razzaque04}, 
GEANT4~\cite{almvz03,razzaque04,McKay_radio} and 
the AIRES+TIERRAS~\cite{TIERRAS,alctz10} code, have yielded results 
compatible to within $\sim 5 \%$. Semi-analytical calculations have also
been performed \cite{buniy02}.
All of these use the same technique to calculate the radio pulse in the
frequency domain, but to our knowledge no full calculation exists in the 
time domain yet. 

All experimental arrangements measure the electric field as a function of 
time, and full understanding of the properties of the pulse as a function of
time is thus also very important. 
Although the conversion from the frequency to the time domain is in principle
straightforward and the algorithm in ZHS computes all required information to
obtain it, there have been a number of doubts concerning the 
unconventional choice of Fourier transform as used in the code~\cite{ZHS92}, as well as the 
sign, phase and causality properties of the pulse \cite{buniy02}, that have complicated the 
analysis and interpretation of data. 

In this article we develop a formalism to
calculate the pulse directly in the time domain. We simultaneously calculate
the pulse of the same electromagnetic shower in both the time and frequency 
domains. An exhaustive comparison yields fully compatible results, makes 
patent the relative advantages of each approach, and sheds new
light into the properties of the radio pulse in the time domain which can be
related to those of the shower and can be of great practical importance in 
interpreting actual data. Some of these properties are discussed in more 
detail suggesting possible applications. 

Although the method developed in~\cite{ZHS92}, and extended here to the
time-domain, has been obtained in the framework of \v Cerenkov radiation,
it derives directly from Maxwell's equations and addresses classical radiation
from charges in a pretty general fashion.
Simple extensions of this work can be used for instance to calculate transition
radiaton as particles cross dielectric media interfaces or to calculate
the complete radiation patterns from charges moving in magnetic fields
including \v Cerenkov radiation, as has been known for long to be important
for ultra high energy air showers.

This paper is structured as follows. 
In Section \ref{theory} we rederive the expression for the electric field in both the
time and frequency domain in a form that can be easily used for practical
applications and make the connection to the expression derived 
in the original ZHS paper \cite{ZHS92}. 
We also discuss some simple current density models and relate them to the results 
of a full electromagnetic shower simulation.
In Section \ref{results} we perform a consistency
check by Fourier-transforming the pulse in time and comparing it to the frequency 
spectrum obtained in the simulations. The summary and outlook constitute the last section. 

%Hadronic and neutrino-induced showers are deferred to a 
%future work when the procedure of calculating pulses
%in the time-domain is implemented in the AIRES+TIERRAS 
%Monte Carlo\cite{TIERRAS,alctz10}. 

%%%%%%%%%%%%%%%%%%%%%%%%%%%%%%%%%%%%%%%%%%
\section{Theory and Monte Carlo implementation} 
\label{theory}

\subsection{Theory}

We start from Maxwell's equations for linear, isotropic, homogeneous and
non dispersive media. In the International System of units: 
%  

%\begin{eqnarray}
\begin{align}
\mathbf{\nabla} \cdot \mathbf{E} &= \frac{\rho}{\epsilon} & 
\mathbf{\nabla} \times \mathbf{E} &= - \frac{\partial \mathbf{B}}{\partial t} \\     
\mathbf{\nabla} \cdot \mathbf{B} &= 0 & 
\mathbf{\nabla} \times \mathbf{B} &= \mu \mathbf{J} + 
                           \mu \epsilon \frac{\partial \mathbf{E}}{\partial t}
%\vec \nabla \cdot \vec E &= {\rho \over \epsilon} & 
%\vec \nabla \times \vec E &= - {\partial \vec B \over \partial t} \\     
%\vec \nabla \cdot \vec B &= 0 & 
%\vec \nabla \times \vec B &= \mu \vec J + 
%                           \mu \epsilon {\partial \vec E \over \partial t}
\end{align}
%\end{eqnarray}
%
where $\rho$ is the charge density of the source, 
$\epsilon=\epsilon_{\rm r} \epsilon_0$ and $\mu=\mu_{\rm r} \mu_0$ are 
the total permittivity and permeability
expressed in terms of the relative ($\mu_{\rm r}$ and $\epsilon_{\rm r}$) and 
free space ($\mu_0$ and $\epsilon_0$) permittivities and permeabilities. 
All effects of induced currents and electric polarization are automatically 
accounted for by the displacement field 
$\mathbf{D}=\epsilon \mathbf{E}$ proportional to
the electric field, $\mathbf{E}$ and the magnetic field strength 
$\mathbf{H}=(\mu)^{-1} \mathbf{B}$, proportional to the 
magnetic field, $\mathbf{B}$. 

We recall the formal solution introducing the vector and scalar potentials
($\mathbf{A}$ and $\phi$): 
\begin{align}
\mathbf{B} &= \mathbf{\nabla} \times \mathbf{A} \\
\mathbf{E} &= -\frac{\partial \mathbf{A}}{\partial t} - \mathbf{\nabla} \phi 
\label{Efield}
\end{align}
that naturally satisfy $\mathbf{\nabla} \cdot \mathbf{B} = 0$, 
and the equation involving the $\mathbf{\nabla} \times \mathbf{E}$ term. Choosing the 
transverse gauge, in which $ \mathbf{\nabla} \cdot \mathbf{A} = 0$, the
two remaining equations imply: 
\begin{align} 
\nabla^2 \phi &= - \frac{\rho}{\epsilon} \\
\nabla^2 \mathbf{A} - \mu \epsilon 
\frac{\partial^2 \mathbf{A}}{\partial^2 t} &=  - \mu \mathbf{J}_\perp 
\end{align} 
where $\mathbf{J}_\perp$ is the transverse current, a divergenceless 
component of 
the current density, which in the limit of observation at large 
distances from the source can be shown to correspond to the projection of the 
current density perpendicular to the direction of observation 
(of unit vector $\hat{\mathbf{u}}$), i.e., 
${\mathbf J}_{\perp}=-\hat{\mathbf{u}}\times(\hat{\mathbf{u}}\times \mathbf{J})$. 

Both equations can be formally solved using Green's functions: 
\begin{align} 
\phi &= \frac{1}{4 \pi \epsilon} \int \frac{\rho(\mathbf{x'},t')}
 {\vert \mathbf{x} -\mathbf{x'} \vert} d^3\mathbf{x'} \label{phisol}\\ 
\mathbf{A} &= \frac{\mu}{4 \pi} \int \frac{\mathbf{J}_\perp(\mathbf{x'},t')} 
{\vert \mathbf{x} -\mathbf{x'} \vert}
\delta \left(\sqrt{\mu \epsilon} \vert \mathbf{x} -\mathbf{x'} \vert- 
         (t-t') \right)
 d^3\mathbf{x'}dt' 
\label{Asol} 
\end{align} 
The first is the familiar solution from electrostatics for the potential 
produced at the position $\mathbf{x}$ by a source
with charge density $\rho(\mathbf{x'},t')$. The second is the
solution of the wave equation with wave velocity 
$(\epsilon_0 \mu_0 \epsilon_{\rm r} \mu_{\rm r}) ^{-{1 \over 2}}$  
smaller than the velocity of light in vacuum, 
$c=(\epsilon_0 \mu_0)^{-{1 \over 2}}$, by a factor 
$n=(\epsilon_{\rm r} \mu_{\rm r})^{-{1 \over 2}}$, the index of 
refraction. The Green's function for the wave equation involves 
a delta function that gives the familiar retarded time, $t'$, 
earlier than the observing time $t$. To evaluate the field at time $t$ at a 
given position $\mathbf{x}$, the current is to be evaluated at a time retarded 
by the time taken by light to reach observation point from point
$\mathbf{x'}$, i.e.  $\vert \mathbf{x}- \mathbf{x'} \vert n/c$.  

\subsection {Radiation from charges traveling in straight lines}
\label{singletrack}

We consider the shower as a superposition of finite particle tracks 
of constant velocity. Each track is completely defined by two limiting times 
$t_1$ and $t_2$, its velocity $\mathbf{v}$ and the position vector of an 
arbitrary point of the track, $\mathbf{x_0}$, which we have chosen 
to correspond to the time $t=0$. The transverse current density entering 
in Eq.~(\ref{Asol}) for a point charge moving with constant 
velocity, $\mathbf{v}$, between the two end points simply 
reads : 
\begin{equation}
\mathbf{J}_\perp(\mathbf{x'},t')=e \mathbf{v}_\perp \delta^3 
\left(\mathbf{x'} - \mathbf{x_0} - \mathbf{v}t' \right) 
\left[ \Theta(t'-t_1) -\Theta(t'-t_2) \right]
\end{equation}
where $-e$ is the charge of an electron, $\mathbf{v}_\perp$ is the projection of the velocity 
onto a plane perpendicular to the direction 
of observation (recall that we consider large distances so that this direction 
is uniquely defined), and $\Theta(x)$ is the Heaviside step function. 

We can now substitute the transverse current into Eq.(\ref{Asol}), integrate 
the three dimensional delta function substituting $\mathbf{x'}$ for 
$\mathbf{x_0}+\mathbf{v}t'$ and approximate the distance between 
$\mathbf{x}$ and $\mathbf{x'}$ by 
$\vert \mathbf{x} -\mathbf{x_0} -\mathbf{v}t' \vert 
\simeq R - \mathbf{v} \cdot \hat{\mathbf{u}} t'$, where 
we define $R=\vert \mathbf{x} - \mathbf{x_0} \vert$. 
In the limit of large distances of observation the denominator 
$\vert \mathbf{x} -\mathbf{x'} \vert$ can be simply 
approximated by $R$. However, we must use the above approximation in the 
argument of the retarding delta function to account for interference effects. 
This corresponds to the Fraunhofer approximation, in which the path difference 
between light pulses emitted at points $\mathbf{x_0}$ and 
$\mathbf{x'}=\mathbf{x_0}+\mathbf{v} t'$ is 
simply the distance between them projected onto the direction of observation. 
As a result the delta function reads $\delta\left(t'(1-n\beta\cos\theta)- 
\left(t-{nR\over c}\right) \right)$, with 
$\mathbf{v}= \mbox{\boldmath{$\beta$}} c$, which can be cast into:
\begin{equation}
{1 \over \vert 1-n \beta\cos\theta \vert} 
\delta\left(t'-{t-{nR\over c} \over 1-n\beta\cos\theta} 
\right) 
\label{delta}
\end{equation}
We note that the recurring 
factor $(1-n\beta \cos \theta)$, with 
$\theta$ the angle between $\mathbf{v}$ and $\hat{\mathbf{u}}$, gives zero for 
the \v Cerenkov angle
$\theta_C$. Moreover the factor changes sign from positive to negative as the 
observation angle changes from being larger to smaller than the \v Cerenkov
angle. 
Now we can perform the integration in $t'$ in Eq.~(\ref{Asol}) which simply implies replacing $t'$
in the step functions by ${t-{nR\over c} \over 1-n\beta\cos\theta}$. 
We now make use of the fact that: 
\begin{equation}
\Theta(ax)=\begin{cases} \Theta(x)~ \text{ if $a>0$},\\
            1 - \Theta(x)~ \text{ if $a<0$} 
\end{cases}
\label{thetacases}
\end{equation}
In this equation we can take $a=(1-n\beta\cos\theta)^{-1}$ and 
$x=t-nR/c-(1-n\beta\cos\theta)t_{1,2}$ 
which allows us to rewrite Eq.(\ref{Asol}) as: 
\begin{equation}
\begin{split}
&\mathbf{A} = {\mu e \over 4 \pi R} \mathbf{v}_\perp \\
& { \Theta(t-{nR \over c} - (1-n\beta\cos\theta)t_1) 
      -\Theta(t-{nR \over c} - (1-n\beta\cos\theta)t_2) 
\over (1-n\beta\cos\theta) }
\label{Atime}
\end{split}
\end{equation}
Note that the modulus in the denominator of Eq.(\ref{delta}) 
is removed because of an effective $\mbox{sgn}(1-n\beta\cos\theta)$ 
that appears when changing the 
argument in the two step functions (according to Eq.~(\ref{thetacases})). 
This expression is easy to implement in a
shower simulation by splitting particle tracks in portions that can be
approximated by uniform motion. 

As $\theta$ approaches the \v Cerenkov angle $\theta_C$ the numerator and 
denominator of Eq.(\ref{Atime}) approach zero. To obtain a formal limit for 
the \v Cerenkov angle we multiply and divide by $\delta t$ 
to obtain:
\begin{equation}
\begin{split}
&R\mathbf{A}(t,\theta)=\frac{e\mu_{r}}{4\pi\epsilon_0
 c^2} {\mathbf{v}}_\perp \delta t % \nonumber
 \\ & %\!\!\!\!\!\!\!\!\!\!\!\!\!\!\!\!\!\!\!\!\!\!\!\!\!\!\!\!\!\!\!\!\!\!\!\!\!\!\!\!\!\!\!\!\!\!\!\!\!\!\!\!\!\!\!\!\!\!\!\!\!\!\!\!\!\!\!\!
\frac{\Theta\left(t-{nR\over c}-(1-n\beta\cos\theta)t_1\right)-
  \Theta\left(t-{nR\over c}-(1-n\beta\cos\theta)t_2\right)}
{(1-n\beta\cos\theta)\delta t}
\label{Atime2}
\end{split}
\end{equation}
The limit $\theta\to\theta_{C}$ is equivalent to $(1-n\beta\cos\theta)\delta t
\to 0$ which can be shown to give the first derivative of the step function,
the function $\delta(t)$. The limit is then: 
\begin{equation}
R\mathbf{A}(t,\theta_C)=\left[\frac{e\mu_{r}}{4\pi\epsilon_0 c^2} \right]
\delta\left(t-{nR\over c}\right) {\mathbf{v}}_\perp \delta t 
\label{AtimeCerLimit}
\end{equation}
We note that the vector potential in this limit is simply proportional (and
parallel) to ${\mathbf{v}}_\perp \delta t$, i.e. to the projection of the 
displacement vector onto a plane perpendicular to the observation direction. 
This expression can also be implemented in a shower simulation provided a 
suitable approximation is made for the delta function.

The expression for the electric field is given by Eq.~(\ref{Efield}) and only
the term with the time derivative of the vector potential gives contribution
to the radiation term so that: 
\begin{equation}
\begin{split}
&R\mathbf{E}(t,\theta)=- \frac{e\mu_{r}}{4\pi\epsilon_0 c^2} {\mathbf{v}}_\perp   %\nonumber
 \\ & %\!\!\!\!\!\!\!\!\!\!\!\!\!\!\!\!\!\!\!\!\!\!\!\!\!\!\!\!\!\!\!\!\!\!\!\!\!\!\!\!\!\!\!\!\!\!\!\!\!\!\!\!\!\!\!\!\!\!\!\! 
\frac{\delta\left(t-{nR\over c}-(1-n\beta\cos\theta)t_1\right)
         -\delta\left(t-{nR\over c}-(1-n\beta\cos\theta)t_2\right)}
{(1-n\beta\cos\theta)}
\label{Etime}
\end{split}
\end{equation}
%\end{align}
%

%
\begin{figure}[tbp]
\centering
\includegraphics[width=9.0cm]{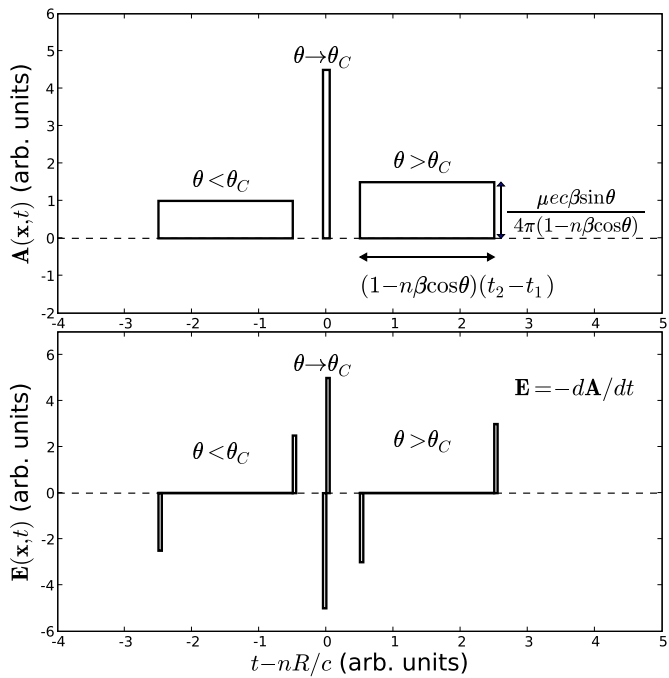}
\caption{Schematic representation of the radiation fields in the time domain induced 
by a single particle with positive charge $e>0$ traveling at constant velocity $\beta$ 
between times $t_1$ and $t_2$. Top panel vector potential (see Eq.(\ref{Atime})). Bottom 
panel electric field (see Eq.(\ref{Etime})). See also text for more details.} 
\label{fig:track_field} % caption for the whole figure
\end{figure}
The radiation field due to a single particle track with $e>0$ is similar to the schematic 
diagram shown in Fig.~\ref{fig:track_field}. 
Such a particle produces radiation when the track starts or ends. 
The two pulses ``as seen" by the observer (placed at angle $\theta$ w.r.t. the particle track) 
are separated by a time interval associated to the difference 
in propagation time $(1-n\beta\cos\theta)\delta t$. 

Let us first
consider an angle exceeding the \v Cerenkov angle so that $(1-n\beta\cos\theta)$
is positive. The electric field of the first pulse corresponds to the start
point of the track ($t_1$) and it is anti-parallel to $\mathbf{v}_\perp$ according
to Eq.(\ref{Etime}), while it is parallel for 
the second pulse which corresponds to the end point ($t_2$). 
The sign of the electric field pulse is opposite to the sign 
of the particle acceleration in both cases. The zero in the shown arrival 
time is arbitrary and corresponds to $t={nR/c}$, i.e. it is a 
reference time associated to the arrival of a signal from the reference position $\mathbf{x_0}$. 
The two pulses associated with the track take place later than this reference time. 

As the angle decreases and becomes smaller than the \v Cerenkov angle, the
situation is reversed: The first pulse corresponds to the end point of the 
track ($t_2$), while the second corresponds to the start point ($t_1$). 
Moreover not only is the arrival of the pulses as seen by the observer
inverted, but both take place before 
the reference time. This apparent acausal behavior is due to the fact that the 
particle travels at a speed greater than that of light in the medium. 
Although the terms responsible for the first and second pulses are interchanged,
and there is a sign change associated with this interchange, it is compensated by
the denominator of Eq.(\ref{Efield}) that also reverses its sign. As a result there is no change in 
the sign of the electric field of the first and second pulses as the 
\v Cerenkov angle is crossed, and the double peak structure at any given time has 
the same qualitative behavior as the observation angle changes. This seems 
physically sound since there can be no discontinuity of the electric field 
across the \v Cerenkov cone boundary. 
 
For observation at the \v Cerenkov angle both signals arrive 
simultaneously. In this limiting case the electric field can be formally 
obtained taking minus the derivative of the delta function given by 
Eq.(\ref{AtimeCerLimit}). This again corresponds to a double pulse first 
antiparallel and then parallel to $\mathbf{v}_\perp$. 

\subsection{Equations in the Frequency Domain}

The expression for the electric field in the frequency 
domain used in the ZHS simulation code 
(Eq.(12) in~\cite{ZHS92}) reads:
\begin{equation}
\mathbf{E}(\omega,\mathbf{x})=
{e \mu_{\rm r}\over 2 \pi \epsilon_0 {\rm c}^2}~
i\omega~{{\rm e}^{i k R } \over R} ~ 
{\rm e}^{i(\omega - \mathbf{k} \cdot \mathbf{v}) {\rm t}_1}~ 
\mathbf{v}_{\perp }~ 
\left[{{\rm e}^{i(\omega - \mathbf{k} \cdot \mathbf{v}) \delta {\rm t}} - 1 
                     \over i (\omega -\mathbf{ k} \cdot \mathbf{v})} \right]
\label{Efreq}
\end{equation}
We recall that this equation has been obtained with the following convention for 
the Fourier transform of the electric field: 
\begin{equation}
\tilde f(\omega)=2\int_{-\infty}^{\infty}{f}(t)~e^{i\omega t}dt
\label{FT}
\end{equation}
where the factor 2 corresponds to an unusual convention (this factor is usually
either 1 or $(2\pi)^{-{1\over2}}$). Applying this Fourier transform definition 
to Eq.(\ref{Etime}) giving the electric field in the time domain we obtain: 
\begin{equation}
\begin{split}
\mathbf{E}(\omega,\mathbf{x})= &
-{e \mu_{\rm r} \over 2 \pi \epsilon_0 {\rm c}^2}~ {1 \over R} ~ 
\mathbf{v}_{\perp} \\ 
&{{\rm e}^{i \omega \left[n R/c + (1-n\beta\cos\theta)\right]{\rm t}_1}- 
{\rm e}^{i \omega \left[n R/c + (1-n\beta\cos\theta)\right]{\rm t}_2} 
\over (1-n\beta\cos\theta)} 
\end{split}
\end{equation}
which can be easily rearranged to give exactly Eq.~(\ref{Efreq}) noting that 
$k={n \omega \over c}$. Moreover if we apply the Fourier transform to Eq.(\ref{AtimeCerLimit}) 
which applies to the limit $\theta\rightarrow\theta_C$ we get: 
\begin{equation}
R \mathbf{A}(\omega,\mathbf{x})=
{e \mu_{\rm r} \over 2 \pi \epsilon_0 {\rm c}^2}~
\mathbf{v}_{\perp} \delta t ~ 
{\rm e}^{i(\omega {\rm t_1} -\mathbf{k} \mathbf{r}_1)} 
~{\rm e}^{ikR} 
\end{equation}
The electric field is obtained taking minus the time derivative which in
Fourier space is just a factor $i \omega$, giving again the same result as Eq.(13) 
in \cite{ZHS92} for the electric field in the frequency domain at 
the \v Cerenkov angle.  

These calculations show the consistency of Eq.(\ref{Etime}) obtained in
the time domain with Eq.(\ref{Efreq}) which gives the field
in the frequency domain: They are simply Fourier transforms of 
each other as expected. 

\subsection{Pulses for Simple Charge Distributions}
\label{simplemodels}

Before performing a Monte Carlo simulation of electromagnetic showers, 
it is interesting to extend the calculations to simple models for the
shower. These models allow us to obtain relations between the shape of the
pulse in the time domain and the time and spatial distribution of the charge. 

A simple yet interesting model consists of a charge $Q(z')$ that rises and 
falls along the shower direction $z'$ and spreads laterally in $x'$ and $y'$. 
Assuming cylindrical symmetry we can write the current associated
to this charge distribution as: 
\begin{equation}
\mathbf{J}(\mathbf{x'},t')=\mathbf{v} f(z',\mathbf{r'})Q(z')\delta(z'-vt')
\label{linecurrent}
\end{equation}
Here $\mathbf{r'}$ is a two dimensional vector in the $(x',y')$ plane transverse
to $z'$, and the 
function $f(z',\mathbf{r'})$ gives the charge distribution in such a plane as 
a function of shower depth, with normalization chosen so that $Q$ indeed gives
the excess charge:
\begin{equation}
\int d^2\mathbf{r} f(z',\mathbf{r})=
            \int_0^{2\pi} d\phi' \int^\infty_0 f(z',r',\phi')=1 
\label{lateralNorm}
\end{equation}
with $\phi'$ the azimuthal angle in cylindrical coordinates.

The simplest case is that of a line current along the $z'$
direction without lateral extension in which $f(x',y')$ is replaced by the two dimensional 
delta function $\delta(x')\delta(y')$. This approximation was also discussed in \cite{alvz99} 
in the frequency domain, where it was referred to as the one-dimensional approximation. 
When such line current is substituted into Eq.(\ref{Asol}) and integrated in 
$x'$, $y'$ and $t'$ making the Fraunhofer approximation, a relatively 
simple expression is obtained that relates the vector potential in the time 
domain to the excess charge $Q(z')$: 
\begin{equation}
\begin{split}
R\mathbf{A} =& \frac{\mu}{4 \pi} \mathbf{v}_\perp \\
& \int_{-\infty}^{\infty} dz' Q(z') 
\delta \left[z'(1-n \beta \cos \theta) - v\left(t-{n R \over c}\right)\right]
\end{split}
\label{Alinecurrent}
\end{equation}
The delta function relates the depth in the shower development $z'$ to the
observation time $t$ through a linear function: 
\begin{equation}
z'=\zeta(t)=\beta{ct-nR \over 1-n\beta \cos \theta}
\label{depthtotime}
\end{equation}
As the observation angle approaches the \v Cerenkov angle, the time 
interval corresponding to the depth spanned by the shower, i.e. the pulse width, 
becomes smaller. We thus recover a familiar result already discussed in 
\cite{ZHS92} although in the frequency domain. 

Performing the integration in Eq.(\ref{Alinecurrent}) yields,
\begin{equation}
R\mathbf{A}=\frac{\mu c \beta}{4\pi}
\frac{\mathbf{v}_{\perp}}
{\vert 1-n\beta\cos\theta\vert} Q(\zeta(t))
\end{equation}
where the delta function in Eq.(\ref{Alinecurrent}) introduces a factor 
$\vert 1-n\beta \cos \theta \vert^{-1}$.

The electric field is obtained taking minus the derivative of the 
vector potential with respect to time: 
\begin{equation}
R\mathbf{E} = - \frac{\mu c \beta}{4 \pi}  
{\mathbf{v}_\perp \over (1-n \beta \cos \theta)\vert 1-n \beta \cos \theta \vert}
\left.{dQ(\zeta) \over d\zeta}\right \vert_{\zeta=\beta{ct-nR \over 1-n\beta \cos \theta}}
\label{Elinecurrent}  
\end{equation}
The factor 
$(1-n\beta\cos\theta)^{-1}$ arises from applying
the chain rule to the derivative of $Q[\zeta(t)]$. 
As a result the pulse in the time
domain can be regarded as the derivative of the development of the charge excess 
along the shower, scaled with the \v Cherenkov factors  
$(1-n \beta \cos \theta)^{-1}$ and $\vert1-n\beta\cos\theta\vert^{-1}$, 
and converted from depth into time through Eq.(\ref{depthtotime}), 
i.e., the pulse is firstly positive and then negative 
with respect to $\mathbf{v}_\perp$ since in a real shower $Q(z')$ corresponds to 
an excess of negative charge. 

A number of interesting results can be directly read off Eq.(\ref{Elinecurrent}). 
If the development curve for the excess charge $Q(z')$ is not symmetric, 
as happens in real showers, the asymmetry in its derivative is directly 
reflected into an asymmetry between the negative and positive parts of the pulse. 
Also it is interesting to note that when the angle of observation is below the 
\v Cerenkov angle, the pulse shape is inverted in time because the early part of
the pulse corresponds to the end of the shower while the beginning of the
shower corresponds to the end part of the pulse, as explained above. Still, 
the polarity of the first and second pulses remains the same because, although 
the slopes before and after shower maximum change sign, there is an extra sign change induced by
the factor $(1-n \beta \cos \theta)^{-1}$. 
This is in complete analogy to what was discussed for a single track. 

In the case of observation in the \v Cerenkov direction, the $z'$ dependence 
of the delta function in Eq.(\ref{Alinecurrent}) disappears and the delta function 
can be factored away from the integral, to give a pulse of amplitude directly 
proportional to the integrated excess track length of the shower. 
The delta function term is due to all parts of the line current being
observed simultaneously at the \v Cerenkov angle.
These are two familiar results already emphasized in \cite{ZHS92}. 

The simulation has shown that the model with the absence of a lateral distribution breaks down at $\vert\theta-\theta_C \vert \lesssim 2.5^\circ$. This result is consistent with that found in \cite{alvz99} where the one dimensional model was studied in the frequency domain.

It is instructive to extend the line current model to a more realistic 
three dimensional current $f(z',r)$ with cylindrical symmetry and current given 
in Eq.(\ref{linecurrent}). In that case the 
expression for the vector potential with two delta functions can be integrated 
in $t'$ and $\phi'$ and the resulting expression involves a double integral 
over cylindrical coordinate $r'$ and the shower depth $z'$:    
\begin{equation}
\begin{split}
&R\mathbf{A}=\mathbf{v_{\perp}}\frac{\mu}{2\pi}
\int_{0}^{\infty} r'dr' \int_{-\infty}^{\infty}dz' f(z',r')Q(z') \\
&\frac{\Theta(n\beta r'\sin\theta-|z'(1-n\beta\cos\theta)-(vt-n\beta R)|)}
{\sqrt{\left[n\beta r'\sin\theta\right]^2-\left[z'\left(1-n\beta\cos\theta\right)-\left(vt-n\beta R\right)\right]^2}}
\end{split}
\label{Alateral}
\end{equation}
This expression, despite being more cumbersome than Eq.(\ref{Alinecurrent}), 
if solved analytically for realistic lateral distribution 
functions, could give insight into useful parametrizations 
of the pulse in the time domain. In any case it can be used 
for numerical simulations. 

In the \v Cerenkov limit Eq.(\ref{Alateral}) becomes
% %
\begin{equation}
\begin{split}
& R\mathbf{A}=\frac{\mu}{2\pi}
\frac{\mathbf{v_{\perp}}}{n\beta \sin\theta_C} 
\int_{-\infty}^{\infty}dz' Q(z') \\
& \int_{\frac {\vert vt-n\beta R \vert}{n\beta\sin\theta_C}}^{\infty}  
\frac{r'dr' f(z',r')}
{\sqrt{r'^2-\left[\frac{vt-n\beta R}{n\beta\sin\theta_C}\right]^2}}
\end{split}
\label{AlateralCerenkov}
\end{equation}
% %
This equation shows that the non-zero width of the electromagnetic 
pulse at the \v Cerenkov angle is the result of the 
lateral distribution of the shower. Although the integral is rather
complicated to evaluate for realistic lateral shower profiles, it can be
shown that for distributions of the form $f(r')=(r')^{-n}$ for 
integers $n>2$ the electric field 
$\mathbf{E}\propto \mathbf{v}_{\perp}\mbox{sgn}(t-nR/c)\vert vt-n\beta R\vert^{-n}$,
which is a fast bi-polar pulse of non-zero width. 

This model still has some limitations. Note that Eq.(\ref{AlateralCerenkov}) predicts a pulse that is symmetric in time while simulations have shown that the pulse at the Cerenkov angle is asymmetric. This is due in part to the radial distribution of velocities of the shower which is not included in the model. The development of a current density vector model that can accurately produce the features of \v Cerenkov radiation is work in progress.

\subsection{Implementation in the ZHS Monte Carlo}

The ZHS Monte Carlo \cite{ZHS92} allows the simulation of electromagnetic
showers and their associated coherent radio emission up to EeV energies \cite{aljpz09}. 
Originally developed in ice \cite{ZHS91}, it has been extended so that electromagnetic 
showers in other dielectric homogeneous media can be simulated \cite{alvz06,aljpz09}. 
The code accounts for bremsstrahlung, pair 
production, and the four interactions responsible for the development of the
excess charge, namely M\o ller, Bhabha, Compton scattering and electron-positron 
annihilation. In addition multiple elastic scattering (according to
Moli\`ere's theory) and continuous ionization losses are also implemented. 
The electron/positron tracks between each interaction are split into subtracks 
so that no subtrack exceeds a maximum depth fixed at 0.1 radiation lengths.
For low energy particles these subdivisions
are actually reduced to ensure that no subtrack is comparable to the particle
range, and they become the step used to evaluate ionization losses 
and multiple elastic scattering. Convergence of results as the step is
reduced has been carefully checked \cite{alvz00}. 

In order to account for interference effects between the radiation emitted due 
to the particles responsible for the excess negative charge, the ZHS code was designed to follow 
all electrons and positrons down to 100 keV kinetic energy threshold, 
as well as to carefully account for time, by considering deviations with respect to 
a plane front moving at the speed of light injected in phase with the primary 
particle. 
In addition to the delays associated to the propagation geometry, 
those due to particles travelling at velocities smaller than the velocity 
of light are accounted for assuming the energy loss is uniform across the 
step. An approximate account is also made of the time delay associated to the 
multiple elastic scattering processes along the step. 

As a result the tracks of all charged particles in a shower are divided into 
multiple subtracks which are assumed to be straight and to have constant velocity. 
The positions of the end points of these subtracks as
well as the corresponding times are readily available by design, 
and they can be used to compute the frequency components of the electric field
making extensive use of Eq.(\ref{Efreq}), taking into 
account the relative phase shift between different tracks because of their 
different starting point positions and time delays. 

In this work we
have extended the Monte Carlo to also calculate the pulse in the time domain. 
A routine has been developed to account for contributions of each 
of these particle subtracks to the vector potential, making 
extensive use of Eq.(\ref{Atime}). Each subtrack contributes a unit 
``rectangle'' to the vector potential, which varies in height,  
``duration'' and sign - see Fig.~\ref{fig:track_field}, depending on the velocity, the relative 
orientation of the track with respect to the direction of observation
and the charge of the particle. When the observation 
direction is very close to the \v Cerenkov angle the delta function in 
Eq.(\ref{AtimeCerLimit}) is replaced by a rectangle corresponding to a nascent 
delta function~\cite{Kelly}. If the sampling time bin width is set to 
$\Delta T$ then a natural choice of nascent delta function is given by
\begin{equation}
\eta_{\Delta T}(t) = 
% \frac{\mbox{rect}(\frac{t}{\Delta t})}{\Delta t} = 
\left\{ 
\begin{array}{l l}
  \frac{1}{\Delta T}, & \quad -\frac{\Delta T}{2} < t \leq \frac{\Delta T}{2}\\
  0, & \quad \mbox{otherwise}\\ \end{array} \right.
% \eta_{\Delta t}(x-x_0)=
\label{eq:discrete_delta_function}
\end{equation}
In this case the base of the rectangle is 
fixed by the intrinsic ``time resolution'' $\Delta T$ of the simulation
and the pulse height depends on $\Delta T$. In practice, the time domain 
radio signal can be reconstructed with an antenna receiver system and 
digital sampling electronics. The time resolution of a single waveform 
is determined by the digital sampling bin width and the high frequency 
cutoff of the receiver system.

Once the vector potential induced by each subtrack 
is defined, the contribution of all charged subtracks 
in the shower is obtained and the vector potential 
is derived with respect to time to obtain the electric 
field in the time domain.  

In the next Section we show 
several examples of the results of this procedure.

%%%%%%%%%%%%%%%%%%%%%%%%%%%%%%%%%%%%%%%%%%

\section{Results}
\label{results}

\begin{figure}[tbp]
\centering
\includegraphics[width=9.0cm]{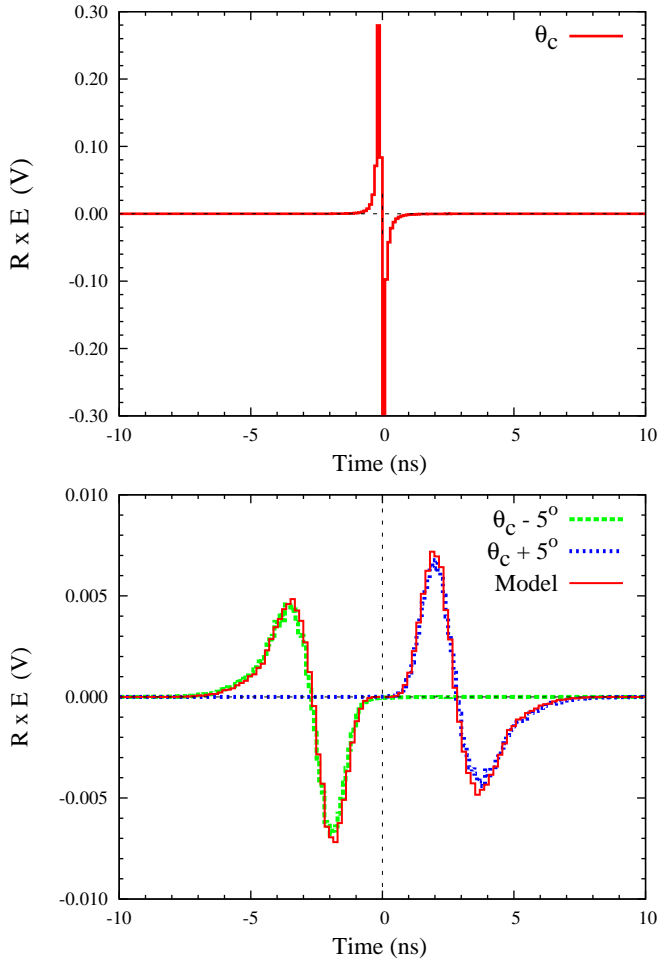}
\caption{Electric field as a function of time as obtained in 
ZHS simulations of a single 1 PeV electron-induced
shower in ice for different observation angles. Top panel: Observation at the \v Cerenkov
angle, bottom panel: observation at $\theta_C-5^\circ$ (long green dashes) and 
at $\theta_C+5^\circ$ (short blue dashes). In the
bottom panel the red solid histograms represent 
the electric field obtained applying Eq.(\ref{Elinecurrent}) to the simulated excess 
negative charge $Q(z)$.} 
\label{fig:efield} 
\end{figure}

In Fig.~\ref{fig:efield} we show the electric field as a function 
of arrival time of the signal obtained with the ZHS code in a single 1 PeV 
electron-induced shower in ice for different observation angles.
The zero in the shown arrival time 
is measured with respect to the arrival time of a pulse 
emitted as the primary particle
initiating the shower is injected in the medium. 

The electric field is parallel to 
the projection of the velocity onto a plane perpendicular to the direction of the observation 
at early times and anti-parallel later on. This is expected after 
the discussion in Section~\ref{singletrack} of the electric field emitted by a single positively
charged particle, 
with the important difference that in a shower the electric field is produced by an excess 
of negative charge and the polarity of the field is reversed with respect to that
shown in Fig.~\ref{fig:track_field}.
Also as in the case of a single track there is no change in the polarity 
of the pulse when observing inside ($\theta<\theta_c$) or outside ($\theta>\theta_c$) 
the \v Cerenkov cone ($\theta_c$). 
The pulse always starts being positive (parallel to $\mathbf v_\perp$) and ends being negative
(antiparallel to $\mathbf v_\perp$) regardless of the observation angle. 
This feature can be used as a discriminator against background events
for neutrino searches. It can 
be also clearly seen that the pulse is broader in time away from the \v Cerenkov cone
than close to it with an apparent duration proportional to $\Delta z \vert1-n\beta\cos\theta\vert/c$
with $\Delta z$ being the spread along the shower axis of the excess charge (see Eq.(\ref{depthtotime})).  
For observation at the \v Cerenkov angle the apparent duration of the pulse
is not zero, despite the fact that the \v Cerenkov factor $\vert1-n\beta\cos\theta_c\vert\rightarrow 0$, 
because the shower spreads out also in the lateral dimensions ($x$ and $y$ directions). 
Also due to our definition of $t=0$ and to the presence of the \v Cerenkov factor
in the $\delta-$functions in Eq.(\ref{Etime}), the pulse occurs at $t>0$ outside
the \v Cerenkov cone and at $t<0$ inside it. 

\begin{figure}[tbp]
\centering
\includegraphics[width=9.0cm]{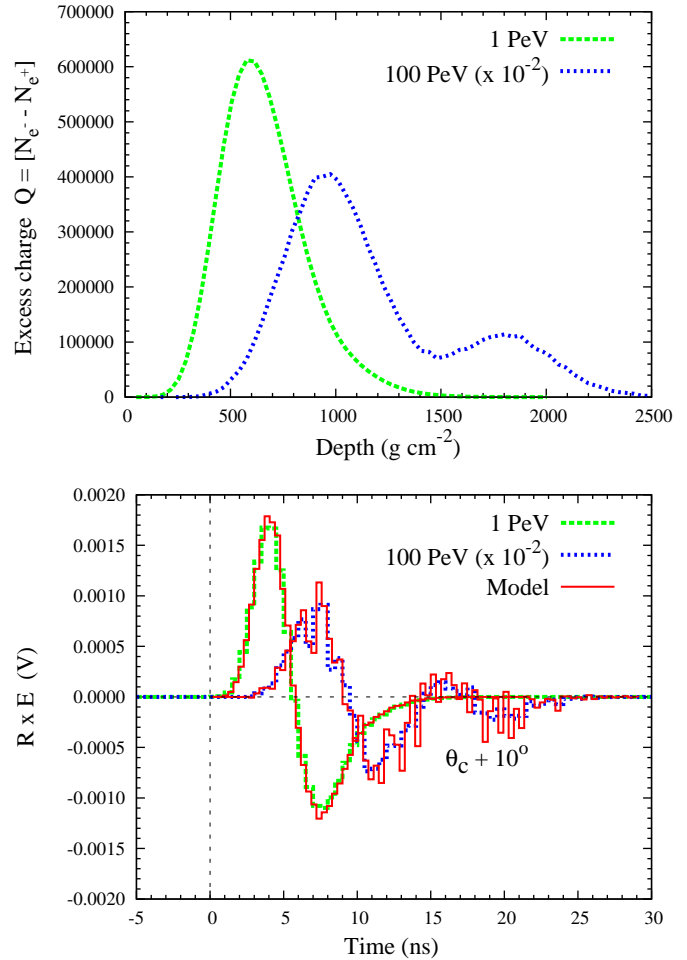}
\caption{Top panel: Longitudinal development of the excess negative charge as obtained
in ZHS simulations of 1 PeV (long green dashes) and 100 PeV (short blue dashes) electron-induced showers
in ice. Bottom panel: Electric field as a function of time generated in the showers shown 
in the top panel (dashed histograms), for observation angle $\theta_C+10^\circ$.
The solid histograms represent 
the electric field obtained applying Eq.(\ref{Elinecurrent}) to the simulated excess charge $Q(z)$
shown in the top panel.} 
\label{fig:efield_lpm} % caption for the whole figure
\end{figure}

According to the simple model developed in Section \ref{simplemodels} the field away from 
the \v Cerenkov angle is proportional to the derivative of the excess charge distribution $Q(z)$ 
with respect to $t$ - Eq.(\ref{Elinecurrent}) - 
or equivalently the derivative with respect to $z$ since there is a linear relation 
between $t$ and $z$ - Eq.(\ref{depthtotime}). 
The ZHS code also gives the longitudinal
profile of the excess charge and we have applied Eq.(\ref{Elinecurrent}) to the 
simulated $Q(z)$, and compared to the electric field obtained directly in the Monte Carlo. 
This is also shown in Fig.~\ref{fig:efield}. The agreement between the electric
field obtained directly in the Monte Carlo simulation (dashed histograms) and what is predicted 
by Eq.(\ref{Elinecurrent}) (solid histograms) is remarkable. The electric field follows the
variation of the excess charge in $z$ or equivalently in $t$. This explains why for 
a fixed observation angle the pulse changes
sign from early to late times (for a typical shower 
$Q(z)$ grows relatively fast, reaches a maximum, and then decreases more slowly 
with depth), 
and why it is asymmetric with respect to the time axis ($Q(z)$ is not a symmetric
function around its maximum). Also when the 
direction of observation is inside the \v Cerenkov cone, the observer sees the derivative
of the beginning of the excess charge distribution first and the corresponding derivative
of the end of $Q(z)$ at later times, while the opposite is true for observations 
outside the \v Cerenkov cone. As a consequence the pulse at $\theta<\theta_c$ looks 
like an antisymmetric copy with respect to $t=0$ of the pulse at $\theta>\theta_c$, 
as can be clearly seen in Fig.~\ref{fig:efield}. An accurate reconstruction of the time 
domain electric field could in principle determine on which side of the Cerenkov cone the event was
observed. On the other hand the shape of the pulse can be conversely used to 
infer the depth development of the shower.

Eq.(\ref{Elinecurrent}) stresses the fact that the features of the excess charge 
distribution are ``mapped" in the time structure of the pulse. 
In particular it is well known that  
electromagnetic showers with energies above the energy scale 
at which the LPM effect \cite{LPM} starts to be effective ($\sim$ PeV in ice \cite{Stanev_LPM}), 
are ``stretched" in the longitudinal dimension and often show peaks in their 
profile \cite{RalstonLPM,Konishi,alz97,klein}. These two features should translate into 
the duration in time also of the pulse and into its time structure that should also exhibit 
multiple peaks. This is shown in Fig.~\ref{fig:efield_lpm} in which due to the LPM effect the longitudinal 
profile of a 100 PeV electron-induced shower exhibits two peaks which appear as 2
positive and 2 negative peaks in the time structure of the pulse. For comparison 
a 1 PeV electron-induced shower not affected by the LPM effect and its corresponding electric field
are also shown. The linear relation between the time domain structure of an electric
field and the shower profile suggests that the longitudinal
profile of the shower could be reconstructed from an observation off
the \v Cerenkov angle.  

\begin{figure}[tbp]
\centering
\includegraphics[width=9.0cm]{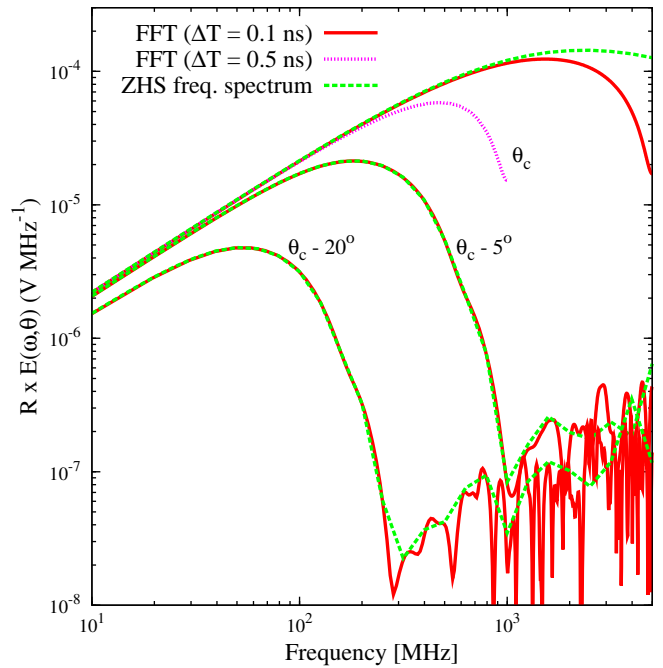}
\caption{Electric field frequency spectrum obtained in 
ZHS simulations of a single 1 PeV electron-induced
shower in ice for different observation angles (green dashed lines). Also 
shown is the 
Fast Fourier Transform (FFT) of the electric field in the time domain 
obtained in simultaneous ZHS simulations of the same shower for two different
time resolutions $\Delta T=0.1$ ns (red solid lines) and $\Delta T=0.5$ ns 
(magenta dotted line - only shown at the \v Cerenkov angle for clarity).} 
\label{fig:efieldFT} 
\end{figure}

The extended ZHS code is able to calculate both the electric field as
a function of time and its Fourier transform from first principles. 
Moreover, the two
calculations can be made simultaneously for the same shower. 
Both calculations can be easily compared by 
performing the Fourier transform of the
pulse calculated in the time domain, following the convention in Eq.(\ref{FT}).
This provides a further check of the two methods, as well as a test of
accuracy in the numerical procedures involved in the calculation of the radio
emission in both domains. An example is shown in Fig.~\ref{fig:efieldFT}, where 
the electric field as a function of frequency as obtained in 
ZHS simulations of a single 1 PeV electron-induced
shower is plotted along with the 
(Fast) Fourier Transform (FFT) of the electric field in the time domain 
obtained in simultaneous ZHS simulations of the same shower.
The agreement between both spectra is very good for frequencies below 
$\omega_{\Delta T} \sim 2\pi/\Delta T$ with $\Delta T$ an arbitrary time
resolution needed for the ZHS simulations 
in the time domain. We do not expect to be able to reproduce the frequency spectrum
at frequencies above $\omega_{\Delta T}$ - proportional to the Nyquist frequency of the system. 
To illustrate this point in Fig.~\ref{fig:efieldFT} we also show 
the Fourier transformed spectrum (at the \v Cerenkov angle) of several time 
domain calculations performed with different time resolutions $\Delta T=0.1$ and 0.5 ns. One can see that the agreement between the frequency spectrum obtained
in ZHS and the Fourier transformed time domain electric field improves as $\Delta T$
decreases as expected. Calculations in the frequency domain are more
advisable near the \v Cerenkov angle.

%%%%%%%%%%%%%%%%%%%%%%%%%%%%%%%%%%%%%%%%%%
\section{Summary and outlook}

In this work we have developed an algorithm to obtain the \v Cerenkov radio 
pulse produced by a single charged particle track in a dielectric medium.
We have implemented this algorithm in the ZHS Monte Carlo with which 
we can predict the \v Cerenkov coherent radio emission emission
of electromagnetic showers in dense dielectric media in 
both the time and frequency domains. 

An observer in the Fraunhofer region, far from the axis of the electromagnetic
shower at an angle $\theta$, sees a bi-polar pulse due to the excess of negative charge 
in the shower. The apparent time duration of the pulse is proportional to $\Delta z~(1-n\cos\theta)/c$
with $\Delta z$ the spread of the shower in the longitudinal direction. At the \v Cerenkov 
angle $(1-n\cos\theta_C)\rightarrow 0$ and the duration of the pulse is mainly determined 
by the lateral extent of the shower. At angles $\theta>\theta_C$, the observer sees first 
the electric field produced by the early stages of the shower, and the field due 
to the end of the shower later on, while the time sequence reverses for observation at $\theta<\theta_C$. 
Regardless of the observation angle, the bulk of the electric field due to the excess negative 
charge is directed along ${\mathbf v}_\perp$ - the projection of the particle velocity onto a 
plane perpendicular to shower axis - at early times and in the opposite direction later on. 
The shape of the pulse maps the variation with depth of the excess charge
in the shower. This information can be of great practical importance for interpreting 
actual data.  

A consistency check performed by Fourier-transforming the pulse
in time and comparing it to the frequency spectrum obtained directly in the simulations
yields, as expected, fully consistent results.
Our results, besides testing algorithms used for reference calculations in the frequency domain,
shed new light into the properties of the radio pulse in the time domain. 

In the future we plan to implement the algorithm for time-domain calculations
of electric field pulses in Monte Carlo simulations of hadronic and neutrino-induced
showers, of great importance for neutrino detectors using the radio \v Cerenkov technique.  
Also we will explore how actual experiments can exploit the richness of information 
contained in the shape in time of the radio pulse to obtain information on the shower
development. This could be of great help in the reconstruction of the parameters of the 
neutrino-induced showers and to discriminate against background events. 

%%%%%%%%%%%%%%%%%%%%%%%%%%%%%%%%%%%%%%%%%%
\section{Acknowledgments}

J.A-M and E.Z. thank Xunta de Galicia (INCITE09 206 336 PR) and 
Conseller\'\i a de Educaci\'on (Grupos de Referencia Competitivos -- 
Consolider Xunta de Galicia 2006/51); Ministerio de Ciencia e Innovaci\'on 
(FPA 2007-65114 and Consolider CPAN) and Feder Funds, Spain.
We thank CESGA (Centro de SuperComputaci\'on de Galicia) for computing resources
and assistance. A. R-W thanks NASA (NESSF Grant NNX07AO05H).
We thank J. Bray and C.W. James for many helpful discussions.


\begin{thebibliography}{99}
%
\bibitem{Askaryan62} G.A.~Askaryan, {\sl Soviet Physics} JETP {\bf 14,2} 
441--443 (1962); {\bf 48 } 988--990 (1965).               
%
\bibitem{ZHS92} E. Zas, F. Halzen, T. Stanev, Phys. Rev. D {\bf 45}, 
362 (1992).
%
\bibitem{allan71}
H.R.~Allan, {\it Progress in Elementary Particles and Cosmic Ray
Physics} {\bf 10}, 171 (1971) (North Holland Publ. Co.), and refs. therein.
%
\bibitem{frichter96} 
G.M. Frichter, J.P. Ralston, D.W. McKay, 
Phys. Rev. D {\bf 53}, 1684 (1996)
%
\bibitem{zheleznykh} R.D. Dagkesamansky and I.M. Zheleznykh, 
%Radioastronomical Method of the Neutrino and Hadron Detection, 
in Proc. of the ICRR International Symposium: Astrophysical Aspects 
of the most energetic Cosmic Rays (Kofu, Japan, November 1990), 
eds. M. Nagano and F. Takahara (World Scientific, 1991) p.373
%
\bibitem{ZHS91} F.~Halzen, E.~Zas, T.~Stanev, Phys. Lett. B {\bf 257}, 
432 (1991). 
%
\bibitem{RICE98}
C. Allen {\it et al.},
New Astron. Revs. {\bf 42}, 319 (1998).
%
\bibitem{Parkes96}
T. Hankins {\it et al.},
Mon.\ Not.\ Royal Astron.\ Soc.\ {\bf 283}, 1027 (1996); 
%
\bibitem{Saltzberg_SLAC_sand} 
D.~Saltzberg {\it et al.}, Phys.\ Rev.\ Lett.\ {\bf 86}, 2802 (2001).
%
\bibitem{Miocinovic_SLAC_sand}The assumption that the current density 
% The fidelity of this one dimensional model begins to break down in region $\vert \theta-\theta_C \vert <2.5^\circ$.

P.~Miocinovic {\it et al.}, Phys. Rev. D {\bf 74}, 043002 (2006).  
%
\bibitem{Gorham_SLAC_salt} 
P.W. Gorham {\it et al.}, Phys. Rev. D {\bf 72}, 023002 (2005). 
%
\bibitem{Gorham_SLAC_ice} 
P.W. Gorham {\it et al.} Phys. Rev. Lett. {\bf 99}, 171101 (2007). 
%
\bibitem{RICE03}
I. Kravchenko {\it et al.}
Astropart. Phys. {\bf 19}, 15 (2003).
%
\bibitem{RICElimits}
I. Kravchenko {\it et al.}
Phys.\ Rev. D \ {\bf 73}, 082002 (2006).
%
\bibitem{Parkes07}
C.W. James {\it et al.} 
Mon.\ Not.\ Royal Astron.\ Soc.\ {\bf 379}, 1037 (2007).
%
\bibitem{ANITAlite}
S.~Barwick {\it et al.}
Phys.\ Rev.\ Lett.\ {\bf 96}, 171101 (2006).
%
\bibitem{ANITA_2009_limits}
P.W.~Gorham {\it et al.}
Phys. Rev. Lett. {\bf 103}, 051103 (2009).
%
\bibitem{ANITAlong}
P.W.~Gorham {\it et al.}
Astropart. Phys. {\bf 32}, 10 (2009)
%
\bibitem{GLUElimits}
P.W.~Gorham {\it et al.}
Phys.\ Rev.\ Lett.\ {\bf 93}, 041101 (2004).
%
\bibitem{Kalyazin}
A.R. Beresnyak {\it et al.}
Astronomy Reports {\bf 49} 2, 127 (2005).
%
\bibitem{NuMoon}
O. Scholten {\it et al.} Astropart.\ Phys.\ {\bf 26}, 219 (2006).
%
\bibitem{LUNASKA} C.W.\ James {\it et al.} submitted
to Phys. Rev. D; C.W.\ James {\it et al.}, Procs.\ $31^{\rm st}$ International
Cosmic Ray Conference, Lodz, Poland (2009), 292.
%
\bibitem{RESUN}
T.R. Jaeger {\it et al.}, arXiv:0910.5949 [astro-ph]
%
\bibitem{ARENA08}
A. Haungs, Nucl. Instr. and Meth. in Phys. Res. A {\bf 604}, S236 (2009),
and refs. therein.
%
\bibitem{alz97} J. Alvarez-Mu\~niz, E. Zas, Phys. Lett. B {\bf 411}, 218 
(1997).
%
\bibitem{alz98} J. Alvarez-Mu\~niz, E. Zas, Phys. Lett. B {\bf 434}, 
396 (1998).
%
\bibitem{alvz99} J. Alvarez-Mu\~niz, R.A. V\'azquez and E. Zas, 
Phys. Rev. D {\bf 61}, 023001 (1999).
%
\bibitem{alvz06}
J.~Alvarez-Mu\~niz, E.~Marqu\'es, R.A.~V\'azquez, E.~Zas,
Phys.\ Rev.\ D {\bf 74}, 023007 (2006).
%
\bibitem{aljpz09} J. Alvarez-Mu\~niz, C.W. James, R.J. Protheroe and E. Zas,
Astropart. Phys. {\bf 32}, 100 (2009).
%
\bibitem{almvz03}
J.~Alvarez-Mu\~niz, E.~Marqu\'es, R.A.~V\'azquez, E.~Zas,
Phys.\ Rev.\ D {\bf 67}, 101303 (2003).
%
\bibitem{razzaque04}
S.~Razzaque {\it et al.} 
Phys.\ Rev.\ D {\bf 69}, 047101 (2004).
%
\bibitem{McKay_radio}
S.~Hussain, D.W.~McKay,
Phys.\ Rev.\ D {\bf 70}, 103003 (2004).
%
\bibitem{TIERRAS}
M. Tueros, S. Sciutto,
Comp. Phys. Comm. {\bf 181}, 380 (2010).
%
\bibitem{alctz10} J. Alvarez-Mu\~niz, W. Rodrigues de Carvalho, M. Tueros and E. Zas,
in preparation (2010).
%
\bibitem{alvz00}
J.~Alvarez-Mu\~niz, R.A.~V\'azquez, E.~Zas,
Phys.\ Rev.\ D {\bf 62}, 063001 (2000).
%
\bibitem{buniy02}
R.V.~Buniy, J.P.~Ralston,
Phys.\ Rev.\ D {\bf 65}, 016003 (2002).
%
\bibitem{Kelly} J.J.~Kelly, 
Graduate Mathematical Physics, Wiley-VCH, Wienheim, (2006).
%
\bibitem{LPM}
 L.~Landau, I.~Pomeranchuk, {\sl Dokl.\ Akad.\
Nauk\ SSSR} {\bf 92}, 535 (1953); {\bf 92}, 735 (1935);
 A.B.~Migdal, Phys.\ Rev.\
{\bf 103}, 1811 (1956);  Zh.\ Eksp.\ Teor.\ Fiz.\
{\bf 32}, 633  (1957) [Sov.\ Phys.\ JETP
{\bf 5}, 527 (1957)].
%
\bibitem{Stanev_LPM}
T.~Stanev {\it et al.},
Phys.\ Rev.\ D {\bf 25}, 1291 (1982).
%
\bibitem{RalstonLPM} J.P.~Ralston, D.W.~McKay, 
Proc.\ of High Energy Gamma-Ray Astronomy Conference (Ann Arbor, Mi 1990), 
ed.\ James Matthews (AIP Conf.\ Proc.\ 220) p.295.
%
\bibitem{Konishi} E.~Konishi, A.~Adachi, N.~Takahashi, A.~Misaki, 
J.\ Phys.\ G: Nucl.\ Part.\ {\bf 17} 719  (1991).
%
\bibitem{klein}
S.R.~Klein, $44^{\rm th}$ Workshop on QCD at Cosmic Energies: 
The Highest Energy Cosmic Rays and QCD, Erice, Italy, 2004.
arXiv:astro-ph/0412546

\end{thebibliography}
\end{document}